\documentclass{article}
\usepackage{spconf,amsmath,graphicx,multirow,tabularx, booktabs,array}
\usepackage{color,soul}
\usepackage{xcolor}
\usepackage{array}

\def\x{{\mathbf x}}

\def\z{{\mathbf z}}

\title{Transformer Transducer: One Model Unifying Streaming and Non-streaming Speech Recognition}

\name{Anshuman Tripathi, Jaeyoung Kim, Qian Zhang, Han Lu, Hasim Sak}
\address{\{anshumant, jaeykim, zhaqian, luha, hasim\}@google.com\\Google Inc., USA}

\begin{document}
\ninept
\maketitle
\begin{abstract}
In this paper we present a Transformer-Transducer model architecture and a training technique to unify streaming and non-streaming speech recognition models into one model.
The model is composed of a stack of transformer layers for audio encoding with no lookahead or right context and an additional stack of transformer layers on top trained with variable right context. In inference time, the context length for the variable context layers can be changed to trade off the latency and the accuracy of the model. We also show that we can run this model in a Y-model architecture with the top layers running in parallel in low latency and high latency modes. This allows us to have streaming speech recognition results with limited latency and delayed speech recognition results with large improvements in accuracy (20\% relative improvement for voice-search task). We show that with limited right context (1-2 seconds of audio) and small additional latency (50-100 milliseconds) at the end of decoding, we can achieve similar accuracy with models using unlimited audio right context. We also present optimizations for audio and label encoders to speed up the inference in streaming and non-streaming speech decoding.

\end{abstract}
%
\section{Introduction}
Past research has shown that having access to future audio context to encode the current audio frame in neural network models significantly improves speech recognition accuracy~\cite{bidirnn,bidilstm,hasim2015,zhang2020transformer}. Bidirectional LSTMs take advantage of future audio context, however the model can only be run when the entire audio is available. In the past few years, models employing self-attention mechanism have achieved state-of-art results for sequence modeling tasks~\cite{vaswani2017attention, dai2019transformer}.
Transformer models encode an input sequence by running self-attention mechanism over left and right context window of each input in sequence. In speech recognition models the future audio context can be encoded by specifying a limited right context~\cite{zhang2020transformer}. Since the right context is limited (unlike bidirectional LSTMs), this allows transformer models with future audio context to recognize speech in streaming fashion with some limited delay. This makes transformer models desirable especially for applications that can afford a higher streaming latency for better recognition quality.

Even in low latency streaming speech recognition systems (e.g. voice-search, dictation), the usability of the system depends on accuracy of the final recognized result. The accuracy of final results can be improved by applying n-best or lattice re-scoring techniques. For longform audio the extra latency introduced by rescoring models is not bounded and the accuracy improvements are also limited due to a lack of diversity in the n-best hypotheses. To address this problem, in this paper, we propose decoding speech using two transformer models in parallel - one with smaller right context for low-latency streaming recognition results and one with larger right context for final results. Unlike rescoring this works well for both short and long utterances. We show that it gives about 20\% relative word error rate improvements over the low-latency results for voice-search task. The extra latency introduced by larger right context decoding is also bounded because of limited right context.

We further show that we can unify low-latency and high-latency models by training a single model that can decode speech in either low latency or high latency mode. We do this by varying right context for transformer layers (variable context layers) during training. By using only final few layers as variable context layers (referred to as Y-model architecture), and applying decoding optimizations presented in this paper, we show that it is possible to do efficient parallel decoding for high latency and low latency modes using one ASR model.

\section{Transformer Transducers}
\subsection{Architecture}
\begin{figure}[h]
\centering
\includegraphics[width=0.45\columnwidth]{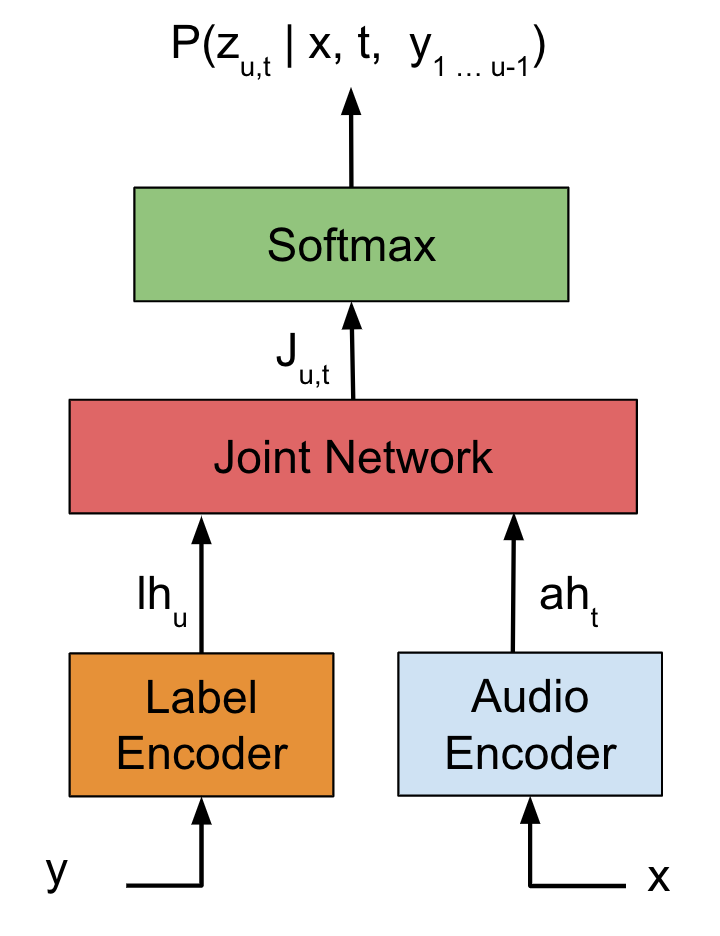}
\caption{Transformer Transducer architecture.}
\label{fig:rnnt_arch}
\end{figure}

\label{sec:tt_arch}
Transformer Transducer~\cite{zhang2020transformer} is a model architecture that can be trained with end-to-end RNN-T loss~\cite{graves:12} using transformer based audio encoder and label encoder. As shown in figure \ref{fig:rnnt_arch},  T-T model predicts a probability distribution over the label space at every time step. 
The probability of an alignment $P(\z|\x)$ can be factorized as
\begin{align}
    P(\z|\x) = \prod_i P(z_i|\x, t_i, \mathrm{Labels}(z_{1:(i-1)})),
\end{align} 
where $\mathrm{Labels}(z_{1:(i-1)})$ is the sequence of non-blank labels in $z_{1:(i-1)}$. 
In T-T architecture $P(\z|\x)$ is parameterized with an audio encoder, a label encoder, and a joint network. 
The model defines $P(z_i|\x, t_i, \mathrm{Labels}(z_{1:(i-1)}))$ as follows:
\begin{equation}
\begin{split}\label{eq:enc-dec}
\mathrm{Joint} = &\mathrm{Linear}(\mathrm{AudioEncoder}_{t_{i}}(\x)) +  \\
        &\mathrm{Linear}(\mathrm{LabelEncoder}(\mathrm{Labels}(z_{1:(i-1)}))))
\end{split}
\end{equation}
\begin{equation}
\begin{split} \label{eq:joint}
P(z_i|\x, t_i, \mathrm{Labels}(&z_{1:(i-1))}) = \\
      &\mathrm{Softmax}(\mathrm{Linear}(\mathrm{tanh}(\mathrm{Joint}))),
\end{split}
\end{equation}
where each $\mathrm{Linear}$ function is a different single-layer feed-forward neural network, $\mathrm{AudioEncoder}_{t_{i}}(\x)$ is the audio encoder output at time $t_i$, and $\mathrm{LabelEncoder}(\mathrm{Labels}(z_{1:(i-1)}))$ is the label encoder output given the previous non-blank label sequence.

\begin{figure}[t]

    \centering
    \includegraphics[width=9cm,height=7cm]{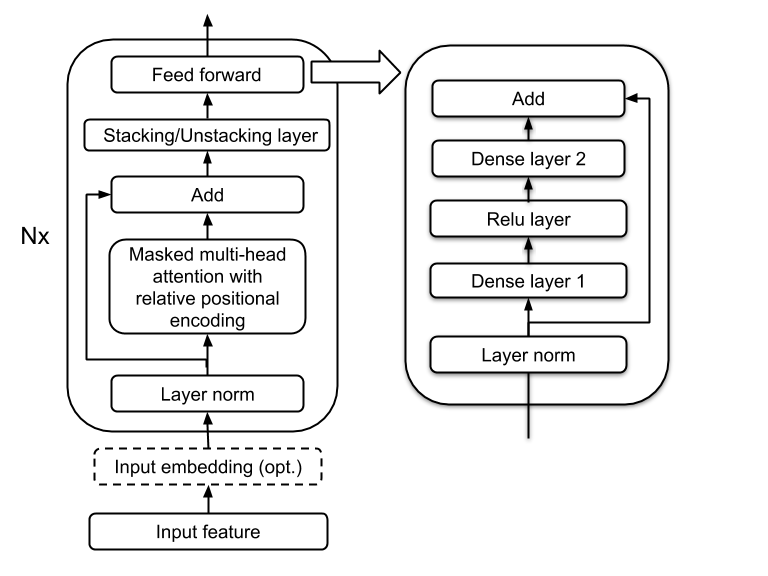}
    \caption{Transformer encoder architecture.}
    \label{fig:transformer_arch}
    \vspace{-11pt} 
\end{figure}

More details about each transformer layer is shown in figure  \ref{fig:transformer_arch}. It contains normalization layer, masked multi-head attention layer with relative position encoding, residual connection, stacking/unstacking layer and feed forward layer. The residual connections are applied with normalized input to the output of the attention layer or feed forward layers. The stacking/unstacking layer can be used to change frame rate for each transformer layer which helps to speed up training and inference. For further optimization, the label encoder can also be a bigram label embedding model. 

\subsection{Variable Context Training}
\label{sec:var_ctx_train}
In self attention block of transformer architecture we compute the self attention over entire input sequence and then mask it based on the left and right context of the layer. In variable context training we keep the left context of the layers constant and sample a right context length from a given distribution. The sampled right context length decides the mask for the self attention. In our experiments we found that randomly sampling right context length for each layer independently led to unstable training, and we had to limit the number of right context configurations. During training we specify right context configuration as a list of right contexts for each transformer layer in the encoder ex: $[0] \times 15 + [8] \times 5$ specifies a right context of 0 for first 15 layers and right context of 8 frames for last five layers. During training we uniformly sample a right context configuration from all these available configurations. We show in section~\ref{sec:y_results} that models trained in this way can be used with any right context configurations used for training.

\subsection{Y-model}
\label{sec:y_arch}
Variable context training allows us to train transformer layers that are capable of using different right context from input, effectively at inference. We apply this technique to train a model that has input followed by several initial layers trained with zero right context and final few layers trained as variable context layers. With this model we can do speech recognition with different future audio context (depending on the right context used at inference time). This can be useful if an application can afford a higher latency for better quality results.
We can also use this model for low latency speech recognition by running two parallel decoders using the same model, one with no or very small right context and one with larger right context. Since we keep the right context of shared layers constant (at zero) we only need to recompute the activation for the final layers with variable context. We call this a Y-model because we have parallel decoding branches running with different right context. We call the decoding branch with smaller right context 'low latency branch' and decoding branch with higher right context 'high latency branch'

\subsubsection{Inference and Recognition Latency}
During inference we stream partial recognition results based on low latency branch (for better model responsiveness) and when the utterance ends we replace the recognition results with the results from high latency branch (for better ASR quality). Depending on the right context used for high latency branch (2.4sec in our experiments), it is always behind the low latency branch in decoding by a finite amount. When the user utterance ends we just need to run high latency branch decoding for the remaining lookahead (e.g. 2.4 secs) and show the final results. Since we have access to all the remaining audio context we can process it very quickly (no need to wait for streaming), by batching all the available context. The additional latency to show these final results at end of the utterance includes run time for computing activation for audio encoder layers with variable context in delayed branch and decoding. In section~\ref{sec:y_results} we show that this can be done very quickly for lookahead audio context of 2.4 seconds, hence resulting in very small additional latency.

\subsection{Constrained Alignment Training}
Transformer-Transducer minimizes RNN-T loss which contains all the alignment paths for each label sequence. Although optimizing label likelihoods should provide advantage on improving ASR performance, a model can have high alignment delay because there's no mechanism to restrict prediction delay in optimizing RNN-T loss. The high alignment delay happens especially for a streaming model which does not have a right context because the model tries to improve prediction accuracy by looking ahead future frames.

Constrained alignment training was originally proposed in~\cite{sak2015acoustic}. It puts constraints on RNN-T loss by masking out high delay alignment paths from reference alignments. To find reference alignments, we used full-attention non-streaming Transformer-Transducer model as a reference alignment model because it showed almost no delay. Based on the reference alignment, we put constrained window on word-boundaries and any word label path outside of the constrained window is masked out from the RNN-T loss. We applied constrained alignment training on Y-model and its evaluation results are in section~\ref{sec: y_constrained_results}.

\section{Inference Optimizations}

In this section, we talk about the optimizations and implementations we did for speeding up the Transformer encoder in streaming and non-streaming applications and the label encoder used for speech decoding. This is required to make parallel decoding feasible for Y-model.

\subsection{Streaming Transformer Audio Encoder}
Streaming audio encoding is the process of taking audio frame(s) and previous states as inputs, and outputting the corresponding encoded feature(s) and next states. Unlike RNN/LSTM based audio encoder, Transformer encoder has no time dependency while encoding audio. In other word, to encode the $i$-th frame, RNN/LSTM based encoders need to first encode from the $0$-th to the $i-1$-th frame before encoding the $i$-th frame, whereas, Transformer encoder can encode all the $i$ frames at the same time in parallel. We refer to the way of encoding a batch of time steps in parallel "batch step" inference. In the later section, we will discuss the effectiveness of batching more time frames in terms of inference speed.

\subsection{Non-streaming Transformer Audio Encoder}

In non-streaming inference, ideally we can run the Transformer encoder exactly the same way it is run during training time. However, in practice, because of the $O(T^2)$ memory consumption in attention matrix computation, the Transformer encoder can be very memory expensive. As showed in~\cite{zhang2020transformer}, limited context Transformer encoder provides about the same quality as infinite context encoder. Based on this model architecture, we can compute the attention matrix for a few $Query$ blocks at a time, and each $Query$ block only attends to a limited set of $Key$s, which makes the memory consumption constant. We refer to this inference method as "query slicing".

\subsection{Decoder Optimizations}
\label{sec:decoder_opt}
In the transformer transducer model, we use a label encoder as an auto-regressive model to encode predicted label history. The label encoder output is combined with the audio encoder output using a dense layer before the softmax function. The decoding speed is highly dependent on the computational complexity of label encoder network because it is run for every hypothesis. For RNN-T models, it has been shown that the label history context length can be reduced significantly without affecting the accuracy of the models~\cite{molimctx}. In this paper, we experimented with two models for label encoding. The first one is a transformer model with limited label context length as described in~\cite{zhang2020transformer}. The other one is an embedding model with a bigram label context. The embedding model learns a weight vector of $d$ dimension for each possible bigram label context, where $d$ is the dimension of audio and label encoder outputs. The total number of parameters is $N^2 * d$ where $N$ is the vocabulary size for the labels. The learned weight vector is simply used as the embedding of bigram label context in the T-T model. Since this is a simple embedding lookup based on label bigrams, the runtime for label encoder is very fast.

To further speed up the inference, we maintain a cache to store and reuse the computed label encoder outputs for the limited label contexts seen so far in decoding since during decoding the model will encode the same label history multiple times for different hypothesis.

\section{Experiments and Results}

\subsection{Data}
For our experiments we use 30K hours of speech data from voice-search application. The test set we use consists of 14K Voice Search utterances with duration less than 5.5 seconds long. The training and test sets are all anonymized and hand-transcribed. The input speech wave-forms are framed using a 32 msec window with 10 msec shift. We use 128 dimension logmel energy features and use as acoustic features after stacking 4 of them and subsampling by a factor of 3, resulting in 30msec features. During training we perform specaugment~\cite{park2019specaugment} on the acoustic features. 

\subsection{Limited Context Decoding}
\label{sec:lim_ctx_results}
Tabel~\ref{tab:label_ctx_wer} shows WER and benchmarking results for different label encoder architectures. From the results we see that there is not much difference in WER when the context is reduce for the label encoder. Similar results have been reported before in~\cite{molimctx}. For the case of limited context of 3 graphemes, we see a huge improvement in speed of decoder over the case when label encoder has 40 labels. This is because of the label encoder output caching as explained in section~\ref{sec:decoder_opt}. For label context 2 it is even faster since the model itself is just a lookup table.
\begin{table}[t]
  \centering
  \begin{tabular}{ccc}
    \toprule
    \multicolumn{1}{c}{\textbf{Label encoder}} & 
    \multicolumn{1}{c}{\textbf{WER}} & 
    \multicolumn{1}{c}{\textbf{RTF}} \\
    \midrule
    40 grapheme context transformer & 4.8 & 0.3\\
    \midrule
    3 grapheme context transformer & 4.8 & 0.02\\
    \midrule
    2 grapheme emb lookup & 4.9 & 0.01 \\
    \midrule
  \end{tabular}
  \caption{WER and RT factors for different label encoders.}
  \label{tab:label_ctx_wer}
\end{table}

\subsection{Y-model results}
We present results on two Y-architecture models:
\subsubsection{Y-Model1}
The audio encoder has 20L of transformers with all the layers trained using variable context with following possible right context configurations: $[0] \times 19 + [4]$, $[2] \times 20$, $[4] \times 20$, and $[32] \times 20$. In all these modes model is trained with an output delay of 4 frames. When evaluating the model with 240ms lookahead we use the right context configuration of $[0] \times 19 + [4]$, when evaluating with 1.2s lookahead we use $[2] \times 20$ and for 2.4 sec look ahead we use the configuration of $[4] \times 20$

\subsubsection{Y-Model2}
The audio encoder has 20L of transformers with first 15 layers trained with no right context and last 5 layers trained with variable context with following possible right context configurations: $[0] \times 19 + [4]$, $[0] \times 15 + [8] \times 5$ and $[0] \times 15 + [16] \times 5$. In all these modes model is trained with an output delay of 4 frames. When evaluating the model with 240ms lookahead we use the right context configuration of $[0] \times 19 + [4]$, when evaluating with 1.2s lookahead we use $[0] \times 15 + [8] \times 5$ and for 2.4 sec look ahead we use the configuration of $[0] \times 15 + [16] \times 5$.

\label{sec:y_results}
Table~\ref{tab:y_model_wer_unconstrained} shows baselines without constrained alignment training. We can see that Y-architecture with a high latency branch of 2.4 sec is very close to the full attention model performance. The same model in low latency mode has much better accuracy than our best streaming model with no right context. We also show the delay in word alignment for all the model evaluations. Since the low latency branch of Y-model2 is still delaying the output by 767 msec, it can get much better performance as it still looks ahead 240 msec + 767 msec by delaying the words.  Another interesting observation is that 2.4 sec lookahead mode for Y model also delays the word alignments, although the full attention model does not delay. This is expected since delaying words helps low latency mode and does not affect high latency mode.
\begin{table}[t]
  \centering
  \begin{tabular}{cccc}
    \toprule
    \multicolumn{1}{c}{\textbf{Model}} & 
    \multicolumn{1}{c}{\textbf{Lookahead}} & 
    \multicolumn{1}{c}{\textbf{WER}}  & 
    \multicolumn{1}{c}{\textbf{Alignment Delay}} \\
    \midrule
    Full context model & 34 sec & 4.8 & 60msec \\
    \midrule
    Left context model & 240 msec & 6.1 & 982msec\\
    \midrule
    \multirow{2}*{Y model1} & 240msec & 6.1 & 883msec\\
                            & 1.2sec & 5.2 & 780msec\\
                            & 2.4sec  & 5.1 & 764msec\\
    \midrule
    \multirow{2}*{Y model2} & 240msec & 5.3 & 767msec\\
                            & 1.2sec & 5.0 & - \\
                            & 2.4sec  & 5.0 & 742msec\\
    \midrule
  \end{tabular}
  \caption{WER with different models and lookaheads.}
  \label{tab:y_model_wer_unconstrained}
\end{table}

Table~\ref{tab:y_model_wer_constrained} shows the results of Y-model with constrained RNN-T loss, we can see that the words are now predicted much earlier, but this leads to degradation in quality of low latency branch. The quality for the higher latency mode is still same since it is not affected by word delays much.

\label{sec: y_constrained_results}
\begin{table}[t]
  \centering
  \begin{tabular}{cccc}
    \toprule
    \multicolumn{1}{c}{\textbf{Model}} & 
    \multicolumn{1}{c}{\textbf{Lookahead}} &
    \multicolumn{1}{c}{\textbf{WER}}  & 
    \multicolumn{1}{c}{\textbf{Alignment Delay}} \\
    \midrule
    \multirow{2}*{Y model2} & 240msec & 6.5 & 119msec\\
                            & 2.4sec  & 4.9 & 74msec\\
    \midrule
  \end{tabular}
  \caption{WER with different models and lookaheads with constrained alignment training.}
  \label{tab:y_model_wer_constrained}
\end{table}
The alignment delay is defined as the mean word alignment difference between reference non-streaming model and the constrained alignment Y model:
\begin{align}
    D = \frac{1}{N}\sum ( T_i^{\textrm{ref}} - T_i^{\textrm{Y}})
\end{align} 
where $T_i^{\textrm{ref}}$ is an alignment time for the $i^{th}$ word from the reference model, $T_i^{\textrm{Y}}$ is an alignment time for the $i^{th}$ word from the Y model and N is the total number of words. For the low-latency mode of Y model, we used extra 4 right context at the last layer to reduce WER loss. Moreover, there is 4 output delay applied to Y model. Therefore, besides alignment delay, Y model has extra 240 msec  for the low latency mode.

Table~\ref{tab:y_model_wer_constrained} shows constrained alignment model. Comparing with unconstrained Y model 2 at Table~\ref{tab:y_model_wer_unconstrained}, alignment delay significantly improved: 767 msec to 119 msec and 742 msec to 74 msec for low and high latency modes, respectively. Moreover, constrained alignment high-latency mode did not show any WER loss because the high-latency mode already has enough right context and restricting its prediction delay did not hurt its performance. However, for the low-latency mode, WER degraded due to the reduced look-ahead frames.

For Y model, more accurate high-latency mode always comes to correct any error from the low-latency mode. Therefore, reducing alignment delay is more important for the low-latency mode, while maintaining high quality performance is the main concern for the high latency mode. The constrained alignment training is well suited for this.
\subsection{Inference Benchmarks}
\begin{figure}[t]
\centering
\includegraphics[width=1.0\columnwidth]{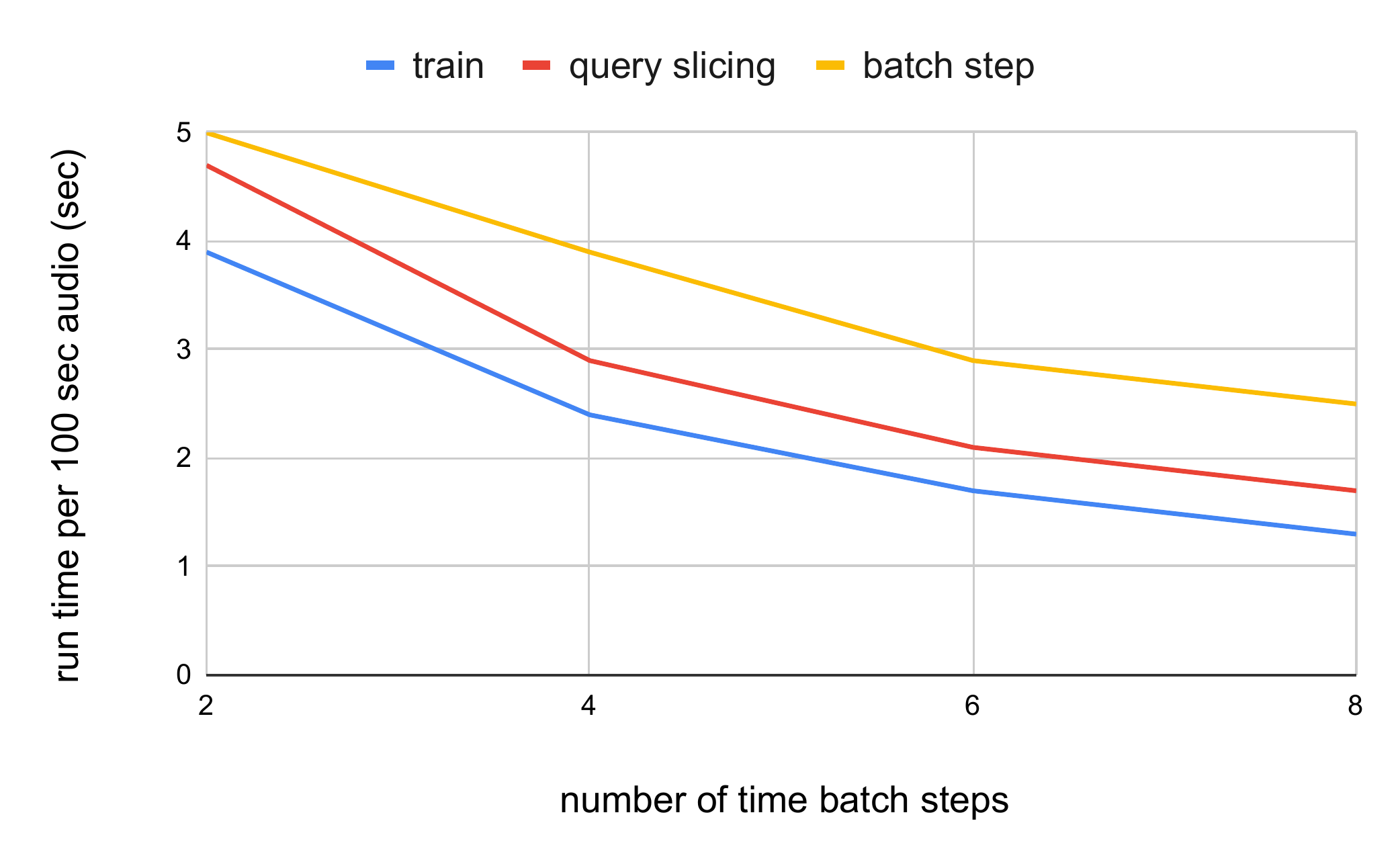}
\caption{Inference speed for batch of 8 utterance with different modes on TPU.}
\label{fig:batch_step}
\end{figure}

In ~Figure\ref{fig:batch_step}, we benchmark the time taken to encode 100 seconds audio with different modes on a single TPU~\cite{googletpu} core with respect to the number of time steps the encoders run at a time. We can see that the inference speed is faster when we encode more time steps at a single inference run for all the modes because it allows better parallelization. We can also see that training mode is faster than query slicing mode, and batch step mode is the slowest among them because of extra overheads for handling the query block, and states. In high latency scenarios, where the encoder encodes 120 frames at a time, the encoding times are even faster. It takes 0.3 second, 0.6 second, and 1.8 seconds to encode 100 seconds audio with training mode, query slicing mode and batch step mode respectively.

\begin{figure}[t]
\centering
\includegraphics[width=1.0\columnwidth]{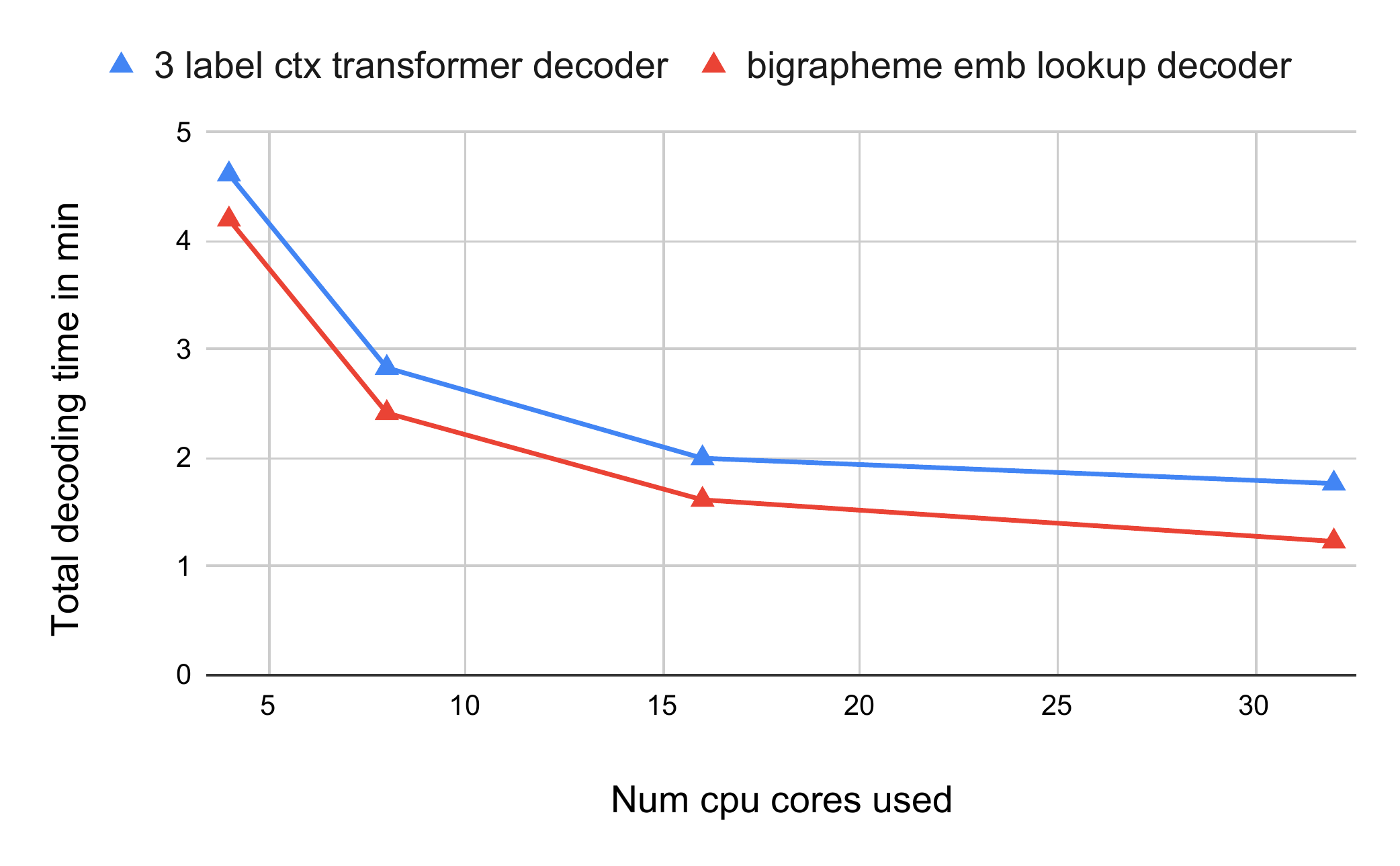}
\caption{Inference speed for query slicing mode on CPU with a 36 minutes audio.}
\label{fig:cpu_query}
\end{figure}

In ~Figure.\ref{fig:cpu_query}, we benchmark query slicing inference mode on desktop CPU for a 36 minutes long audio. We can see that this audio can be recognized in less than 3 minutes with 8 CPU cores (i.e. less than 8\% real time factor). The recognition time can be further improved with a bigrapheme embedding lookup decoder with slight WER regression.

\section{Conclusions}
\label{sec:conclusions}
We describe a method to train a Transformer-Transducer model that allows training a single model that can decode speech in both low latency and high latency modes. We also propose Y-model trained with variable right context at penultimate layers as an efficient solution for unifying low-latency and high-latency decoding modes. The low latency mode can be used for streaming recognition results while the final recognition results are obtained from high-latency mode. We show that with a small future audio context of 2.4 sec, we can get very similar accuracy to a model with full audio context. We also show that the extra latency at the end of decoding from high latency mode can be made very small with parallel encoding of audio frames and decoder optimizations.

\newpage
\bibliographystyle{IEEEbib}
\bibliography{main}

\end{document}